 \documentclass[authoryear,preprint,review,12pt]{elsarticle}

\usepackage{amssymb}

\def\astrobj#1{#1}
\journal{New Astronomy}

\begin{document}

\begin{frontmatter}



\title{An asteroseismic study of the Southern $\beta$ Cephei star \astrobj{ALS 3721}}


\author[l1]{C. Ulusoy}
\author[l2]{E. Niemczura}
\author[l3]{B. Ula{\c s}}
\author[l1]{T. G{\"u}lmez}
\address[l1]{Department of Physics, University of Johannesburg, P.O. Box 524, APK Campus, 2006, Johannesburg, South Africa}
\address[l2]{Astronomical Institute of Wroclaw University, ul. Kopernika 11, 51-622 Wroclaw, Poland}
\address[l3]{Department of Astronomy and Space Sciences, University of Ege, 35100, Bornova, {\.I}zmir, Turkey}
\begin{abstract}
We present the results of a new investigation aimed to identify the pulsational characteristics of the Southern $\beta$ Cephei star ALS 3721. Spectroscopic and multicolour photometric data were acquired at the South African Astronomical Observatory (SAAO), South Africa in 2011. Frequency analysis showed that the oscillations of ALS 3721 could be attributed to the two main frequencies with higher significance. Stellar parameters and projected rotational velocity obtained by the spectra were also used to perform photometric mode identification. In order to  determine spherical harmonic degrees ($l$), a principal method was followed by comparing the observed light amplitude ratios in different passbands with those computed from non--adiabatic pulsation models. In general case, therefore, the spherical harmonic degrees corresponding of the frequencies were found in the expected $\beta$ Cephei range.
\end{abstract}


\end{frontmatter}

\section{Introduction}

The $\beta$ Cephei stars are defined as a group of early B type near main sequence pulsators  (B0.5-B2.5) with the masses between about 8 and 18 $M_\odot$  \citep{sta05}. They show both mono and multiperiodic short term pulsational variability in the periods of several hours.  Although their general pulsational behaviour is considered by low order $p$ (pressure) modes, a number of them pulsate in low order $g$ (gravity) modes and pulsations are driven by $\kappa$ mechanism through iron group elements in the partial ionization layers of their stellar structure \citep{cox92,dzi93a,dzi93b}. Like most other type of pulsating stars, $\beta$ Cephei stars are represented with a well known instability strip in the Hertzprung-Russel Diagram (HRD). However, the blue part of this instability domain is not well populated since there is lack of chance to detect unstable modes for such massive ($\geq$ 25 $M_\odot$) stars with strong stellar winds \citep{pam99}. In a similar way, it is also very difficult to observe very low amplitude modes at the red part of the strip from the ground based observations. A detailed review of the  {\it Kepler} observations of B type variables were recently reported by \cite{bal11}. They pointed out that instability domain among B type pulsating stars may also be divided into two subgroups in the HRD as the metal opacity bump occurred in different layers for each star group: $i)$ high-luminosity $\beta$ Cephei and $ii)$ low-luminosity SPB (Slowly Pulsating B stars). In this case, the most recent asteroseismic studies for these type of variables are being more important to understand their strange physical structures.

The star \astrobj{ALS 3721} is a member of the Southern open cluster \astrobj{NGC 6200} located in the constellation Ara. The first observations of the cluster was made by \cite{fit77}. The authors  derived the colour excess \($E(B-V)$=0^{m}.63 \pm 0^{m}.07 \) and the distance modulus \(11^{m}.9 \pm 0^{m}.2 \) of the cluster. \astrobj{ALS 3721}, on the other hand, was classified as B0.5III by \cite{whi63} and B1III by \cite{fit77}. \cite{pig05} indicated that the star has $\beta$ Cephei  type oscillation characteristics based on the ASAS-3 survey. \cite{pig08} analyzed the ASAS-3 data in order to determine the oscillation frequencies of \astrobj{ALS 3721}, and they reported the three independent frequencies with highest amplitude. Recently, \cite{ulu12} analyzed the new photometric data taken from the SAAO by using the Lomb-Scargle periodograms and they referred to ten oscillation frequencies that are reached in the $B$ light curve.

In this paper, we present a detailed study of the pulsation properties of \astrobj{ALS 3721}. Spectroscopic observations and analysis of the target are explained in the next section. In the third section the details of the photometric observations, reduction and frequency analysis are given while the mode identification for each detected frequency, which is the main goal of this study, and its results are stated in the fourth section. The study is summarized and results are discussed in the last section.

\section{Spectroscopy}

The spectroscopic data were used to derive the fundamental stellar atmospheric parameters and the rotational velocity, $v \sin i$. The observations were carried out on 18 June 2011 with the Robert Stobbie Spectrograph (RSS) longslit instrument mounted at the Southern African Large Telescope (SALT), Sutherland, South Africa. The specifications for the RSS instrument yielded $R\sim 3000$ spectral resolution of given characteristics related to the spectra taken for \astrobj{ALS 3721}. Four spectra were obtained with exposure times set to 60 s and signal-to-noise ratios were achieved from 80 to 100  with a spectral range of 3925 and 5990 \AA. Reductions for the RSS spectra were performed using the {\tt IRAF-ONEDSPEC} and the {\tt TWODSPEC} packages, also the normalization of spectra were carried out by means of the {\tt CONTINIUUM} package.
In order to obtain atmospheric parameters of \astrobj{ALS 3721}  we compared the low-resolution averaged SALT spectrum with theoretical spectra from the {\tt TLUSTY/SYNSPEC} codes given in the {\tt BSTAR2006} grid \citep{lan07}. The {\tt BSTAR2006} grid consists of 1540 metal line-blanketed, non-LTE, plane-parallel, hydrostatic model atmospheres. The synthetic spectra were calculated for effective temperatures between 15000 and 30000 K with a step equal to 1000 K and surface gravities ranging from 1.75 to 4.75 dex with a step equal to 0.25 dex. In our fitting we adopted fluxes calculated for solar metallicity. The spectra are available for microturbulence velocity equal to 2 km\,s$^{-1}$. The model spectra were convolved with rotational and instrumental profiles.

\begin{figure}
\includegraphics[scale=0.7]{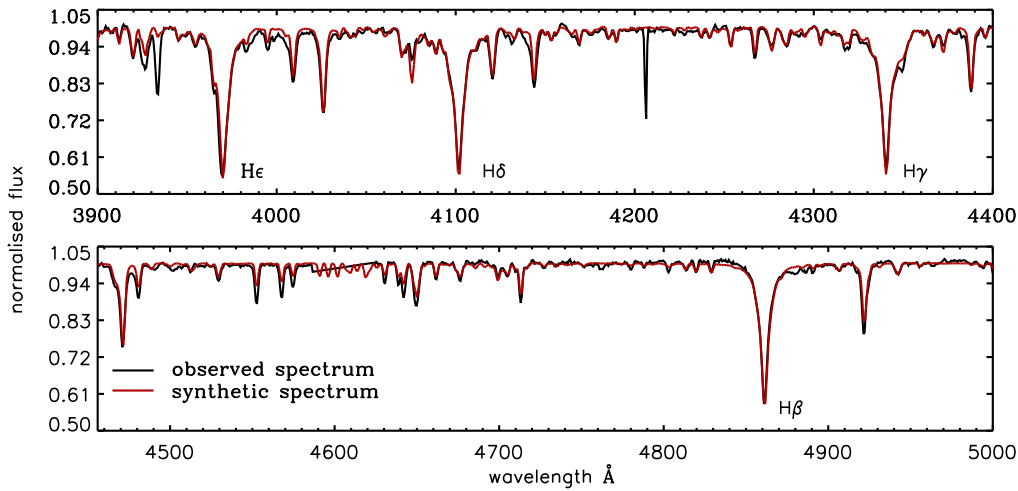}
\caption{Observed (black) and synthetic (red-fitted) spectra of ALS 3721 for selected region.The  observed lines fitted to model fluxes with $T_{\rm eff} = 24000\pm1000$\,K and $\log g = 3.5\pm0.1$.}
\label{figs1}
\end{figure}

The determination of surface gravity $\log g$ and effective temperature $T_{\rm eff}$ was based on the whole available spectrum. The important diagnostic features are clearly visible. Four Balmer lines (H$\epsilon$, H$\delta$, H$\gamma$ and H$\beta$), He\,I lines (4026, 4388, 4471, 4713, 4921 and 5876\,\AA), silicon lines Si\,III (4552, 4568, 4575\,\AA) and blended lines of other elements, like Si\,II, O\,II, C\,II, N\,II, S\,II and Fe\,III were identified. The best fit of the model fluxes to the observed lines was obtained as a result of minimizing residual standard deviation of the fit (Fig.~\ref{figs1}). The analysis resulted in $T_{\rm eff} = 24000\pm1000$\,K and $\log g = 3.5\pm0.1$.  Additionally, the projected rotational velocity $v \sin i$, was derived to be equal to $100 \pm 20$\,km\,s$^{-1}$ and it is obvious that higher microturbulence velocity is needed.

\section{Photometry}

\begin{figure}
\begin{center}
\includegraphics[scale=0.7]{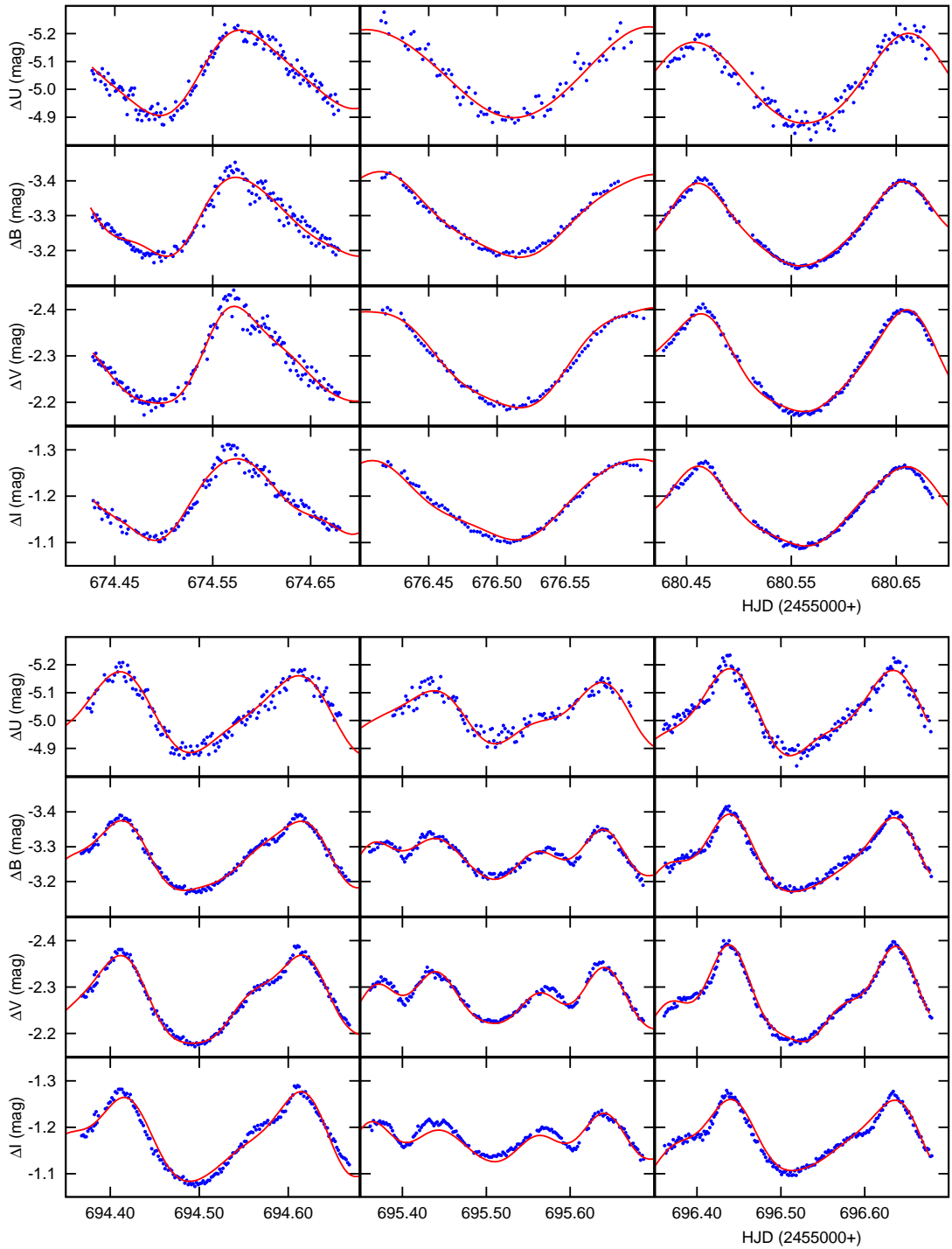}
\caption{Observed (dotted) and theoretical (solid) $UBVI$ light curves of  \astrobj{ALS 3721} for six different observing nights.}
\label{figp1}
\end{center}
\end{figure}

The multicolour photometric data were used to calculate the amplitude-colour relation for the two strongest frequencies  of \astrobj{ALS 3721}. This was achieved using the data taken by \cite{ulu12}.  They carried out the observations with the 75-cm and the 100-cm telescopes at the Sutherland station of the SAAO. Since the data obtained from the 75-cm telescope is scattered than expected and a large time gap between two observing runs from the different instruments exist we included only data from the 100-cm Elizabeth Cassegrain telescope to our investigation. The details of the study are explained by \cite{ulu12}. During the observations, a total of 780, 866, 871, and 853 observing points in the Johnson {\it{U}}, {\it{B}}, {\it{V}} and {\it{I}} filters were collected, respectively. Data reduction was performed following standard {\tt IRAF} routines. In the pre-reduction process, the bias and the dark frames were subtracted, the flat-field correction was applied to each of CCD images. Instrumental magnitudes for all stars in the field of view were measured by aperture photometry technique. All of the images were processed using the {\tt DAOPHOT II} package \citep{ste87} which is also included in {\tt IRAF} following the standard tasks of the package. The data were reduced relative to only one comparison star, GSC 8330 2327, due to showing the least variability in the observed CCD field. The observed light curves of the star in the $UBVI$ filters can be seen in Fig.~\ref{figp1}.

Oscillation frequencies were computed with the {\tt PERIOD04} \citep{l05} software for each filter. The sofware package is based on the classical Fourier analysis and applies least--squares tecnique to input data by the fitting formula:

\begin{equation}
\label{e1}
f\left( t \right) = Z +  \displaystyle\sum\limits_{i}^n A_{i}\sin \left( 2\pi\left( \Omega_{i} t + \Phi_{i}  \right) \right)
\end{equation}
where $Z$ is the zero level of magnitudes, $A_i$ is the amplitude of corresponding frequency, $\Omega_{i}$ is the frequency value, $t$ is time and $\Phi_{i}$ is the phase of given frequency. The signal-to-noise ratio $S/N > 4$ threshold was adopted as a criterion following \cite{bre00}. The theoretical frequency resolution was found to be $1/T = 0.4494$. Since $\beta$ Cephei type oscillation frequencies are generally detected in the range of 4-20 d$^{-1}$ \citep{bal11} we restricted our analysis on the frequency range between 0 d$^{-1}$ to 15 d$^{-1}$.

\begin{figure}
\begin{center}
\includegraphics{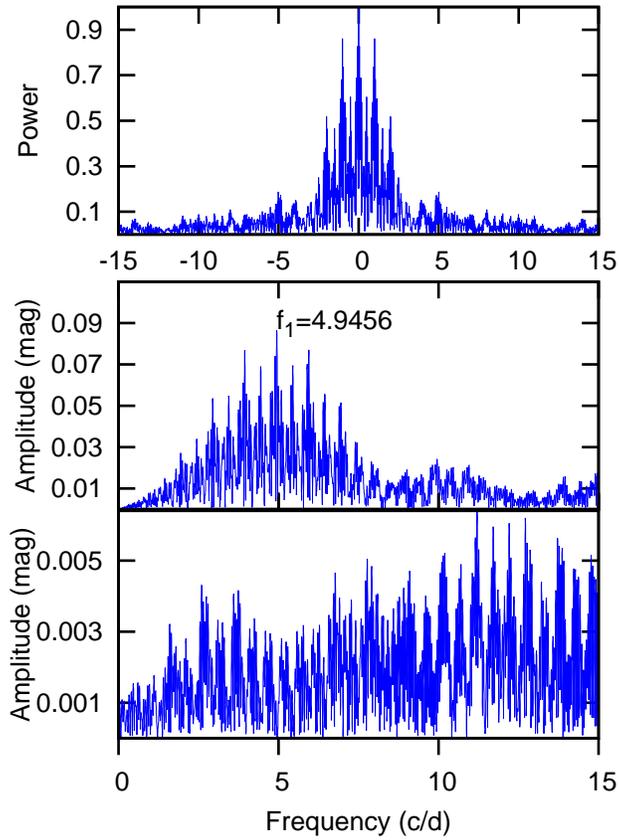}
\caption{The frequency spectra of \astrobj{ALS 3721} in the $B$ filter. The top panel shows the spectral window. The middle panel represents the periodogram yielded before the analysis while the bottom one shows after pre--whitened all listed frequencies in Table~\ref{tabp1}.}
\label{figp2}
\end{center}
\end{figure}

The analysis completed with the detection of the independent and combination frequencies. The results correspond that two frequencies with the highest amplitudes can be assigned for the oscillation of \astrobj{ALS 3721}, $f_{1}$=4.9457, $f_{2}$=5.3928 with the amplitudes of $A_{1}$=0.08404, $A_{2}$=0.02517, respectively. The amplitude spectra of $B$--filter data before prewhitening of any frequency between mentioned interval is shown in the middle panel of Fig.~\ref{figp2}. Peaks with higher amplitude around 5 d$^{-1}$ is clearly seen from the figure. The bottom panel represents the final spectra after overall process. The obtained frequencies, amplitudes, phases and S/N values for all light curves after the analysis  are sorted in Table~\ref{tabp1}. Signal--to--noise ratios were computed in the interval of 2 d$^{-1}$. Light variation of the star in the $UBVI$ filters corresponding of  multiperiodic oscillations is shown for six different nights in Fig.~\ref{figp1}. Solid lines in the figure show the agreement between analysis results and the observational data.
\begin{table}
\caption{Computed frequencies, amplitudes and phase shifts during the solution. Signal-to-noise ratio per frequency is given in the last column. The standard deviations in the last digits are given in parentheses.}
\label{tabp1}
\tiny
\begin{tabular}{lrrrr}
\hline
ID	  &   \multicolumn{1}{c}{$f$ (d$^{-1}$)}&    \multicolumn{1}{c}{A (mag)} &   \multicolumn{1}{c}{$\phi$}	 &  \multicolumn{1}{c}{$S/N$} \\
\hline
& &\multicolumn{1}{c}{$U$} & &\\
$f_{1}$&4.94568(25)&0.12067(124)&0.8872(16) &65.0 \\
$f_{2}$&0.0719(13) &0.02408(124)&0.7860(82)&10.7 \\
$f_{3}$&5.3951(10) &0.02965(124)&0.9077(67)&14.8 \\
$f_{4} \approx 2f_{1}$&9.8727(12) &0.02616(124)&0.2369(77)&14.4 \\
$f_{5}$&10.4076(24) &0.01216(124)&0.9613(157)&7.2 \\
$f_{6}$&3.9682(10) &0.01118(124)&0.9042(166)&5.7 \\
$f_{7}$&11.0997(33) &0.00851(124)&0.3956(216)&4.9 \\
& &\multicolumn{1}{c}{$B$} & &\\
$f_{1}$&4.94565(8)&0.08404(55)&0.3131(5) &54.6 \\
$f_{2}$&5.3928(6) &0.02517(55)&0.1529(36)&15.6 \\
$f_{3}$&0.0719(9) &0.01521(55)&0.9118(60)&20.8 \\
$f_{4} \approx f_{3}+2f_{2}$&10.8750(6) &0.02039(55)&0.9725(44)&24.7 \\
$f_{5} \approx f_{1}+f_{4}-f_{2}$&10.4143(11) &0.01206(55)&0.0711(33)&12.7 \\
$f_{6} \approx f_{1}$&4.9726(1) &0.01236(55)&0.4157(9)&8.2 \\
$f_{7}$&13.8882(13) &0.01095(55)&0.9684(82)&4.8 \\
$f_{8}$&1.3707(15) &0.00806(55)&0.4107(99)&8.1 \\
$f_{9} \approx 3f_{2}-f_{1}$&11.2030(10) &0.00826(55)&0.3873(102)&10.2 \\
& &\multicolumn{1}{c}{$V$} & &\\
$f_{1}$&4.94563(14)&0.07809(44)&0.0632(9) &85.5 \\
$f_{2}$&0.02472(11)&0.02607(44)&0.1498(7) &18.6 \\
$f_{3}$&5.3905(5)&0.02224(44)&0.7944(32) &23.9 \\
$f_{4} \approx 2f_{1}$&9.8706(6)&0.01927(44)&0.5048(37) &21.6 \\
$f_{5} \approx f_{1}+f_{3}+2f_{2}$&10.4121(12)&0.00929(44)&0.3159(77) &11.3 \\
$f_{6} \approx f_{1}$&4.97259(9)&0.01297(44)&0.3456(6) &14.2 \\
$f_{7} \approx 3f_{6}$&14.8993(10)&0.01055(44)&0.2958(69) &8.1 \\
$f_{8}$&1.3167(15)&0.00657(44)&0.9132(94) &5.2 \\
$f_{9}$&11.1514(14)&0.00791(44)&0.3691(90) &10.7 \\
$f_{10} \approx f_{1}+2f_{4}-f_{5}$&14.2365(19)&0.00531(44)&0.3175(121) &4.7 \\
$f_{11}$&3.5413(34)&0.00411(44)&0.9863(220) &4.8 \\
& &\multicolumn{1}{c}{$I$} & &\\
$f_{1}$&4.94338(22)&0.05811(53)&0.8090(14) &51.0 \\
$f_{2}$&5.39502(54)&0.02422(53)&0.7542(35) &19.8 \\
$f_{3}$&0.0719(10)&0.01366(53)&0.7355(62) &5.6 \\
$f_{4} \approx f_{3}+2f_{2}$&10.8727(7)&0.01924(53)&0.1160(44) &9.0 \\
$f_{5} \approx f_{1}+f_{4}-f_{2}$&10.4143(10)&0.01209(53)&0.2394(70) &6.7 \\
$f_{6}$&4.1592(13)&0.01001(53)&0.3548(84) &6.4 \\
\hline
\end{tabular}
\end{table}

We conclude that \astrobj{ALS 3721} oscillates in the frequency range of $\beta$ Cep type oscillators which were also confirmed by \cite{pig08} previously using ASAS-3 data. Table~\ref{tabp2} compares our analysis results of $V$ light curve to that of determined by \cite{pig08}.

\begin{table}
\caption{Comparison of Fourier analysis results of $V$ light curve to PP08 which stands for \cite{pig08}.}
\label{tabp2}
\scriptsize
\begin{tabular}{lrr}
\hline
ID	  & \multicolumn{1}{c}{This study}  &     \multicolumn{1}{c}{PP08} \\
\hline
$f_{1}$(c/d)&4.9456&4.9488\\
$A_{1}$(mag)&0.0781&0.0872\\
\\
$f_{2}$(c/d)&\multicolumn{1}{c}{--}&4.9246\\
$A_{2}$(mag)&\multicolumn{1}{c}{--}&0.0151\\
\\
$f_{3}$(c/d)&5.3905&5.3989\\
$A_{3}$(mag)&0.0261&0.0116\\
\\
$f_{4}^{*} \approx f_{1}+f_{2}$(c/d)&9.8706&9.8734\\
$A_{3}$(mag)&0.0193&0.0105\\
\hline
\end{tabular}
\\
$^*$ The combination frequency contains $f_{2}$ which we could \\
not obtain but published by PP08 previously.
\end{table}

\section{Mode Identification}
There are different mode identification methods that can be used to identify the pulsation mode geometry of a star. An oscillation mode is specified by three spherical harmonic quantum numbers; $l$ the degree of the pulsation mode, $m$ the azimuthal order, and $n$ the radial overtone. In the photometric case, spherical harmonic degree, $l$ can be derived by comparing observed amplitude ratios, or phase differences for a given passband with the predicted ones. Hence, mode identification from multicolour photometry is an essential requirement for modelling the oscillations in pulsating stars. For our study, we used the {\tt FAMIAS} software package \citep{zim08} to determine the wavelength dependence of the photometric amplitudes corresponding to the mode degrees of the first two strongest frequencies found in \astrobj{ALS 3721}. In {\tt FAMIAS}, the photometric module applies the aforementioned method with precomputed model grids of non--adiabatic stellar models for a certain range of stellar parameters  in a number of photometric passbands including the Geneva, Johnson/Cousins and Str\"{o}mgren systems. Whereas the determination of the $l$ degrees strongly depends on pulsational input parameters, our calculations are performed in a wide range of the model grids to fit the best estimation of the $l$ degrees.

\begin{figure*}
\includegraphics[scale=0.8]{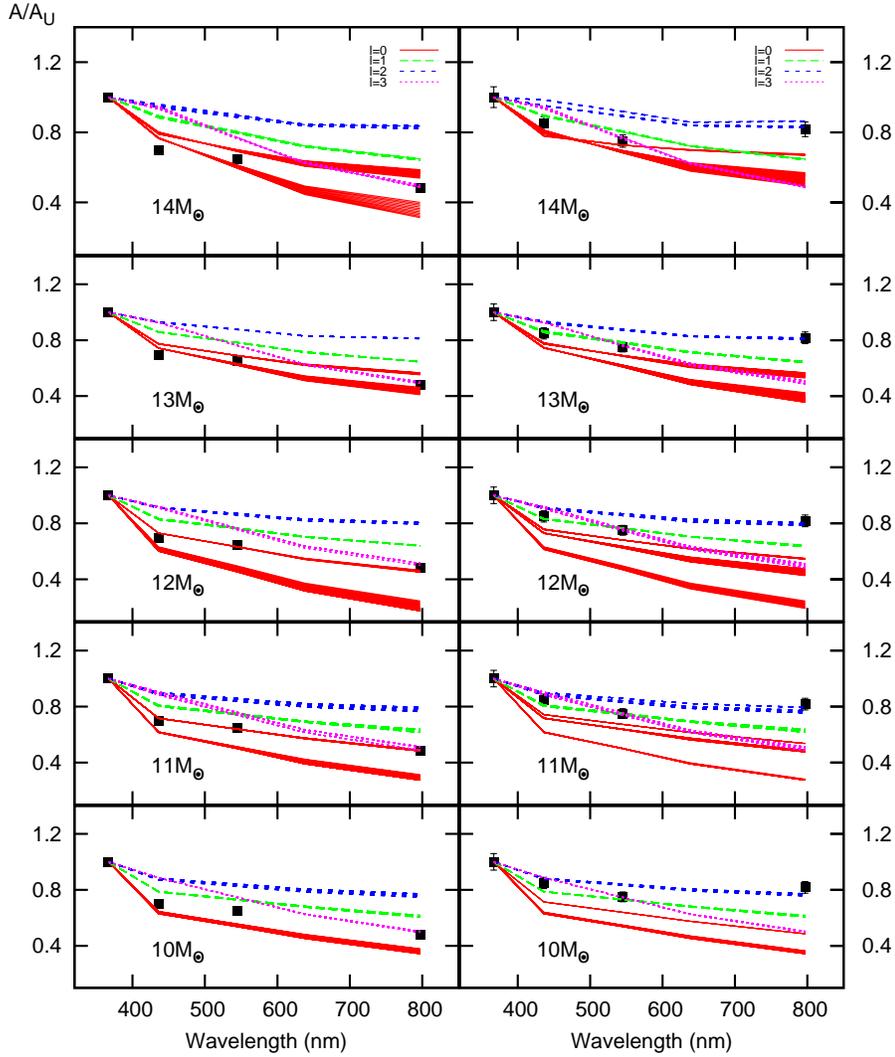}   \\
\caption{Observed and theoretical $UBVI$ amplitude ratios for \astrobj{ALS 3721}. The colour amplitudes are given as a function of the central wavelength normalized at the the U filter for $f_1$ (left) and  $f_2$ (right). The coloured lines in the panels represent different spherical harmonic degrees corresponding to different stellar models  in the range of  $23000 < T_{\rm eff} < 25000$~K and $T_{\rm eff}$, and $\log g$ $3.2 < \log g < 3.8$. The red (solid) lines indicate $l=0$ , the green (long-dashed) $l=1$, the dark blue (dashed) $l=2$ the purple (dotted) $l=3$ . The models computed a mass range of between 9 and 15 M$_\odot$. The black crosses represent the observed values with their 1$sigma$ standard deviations for $f_1$ and  $f_2$.}
\label{figm1}
\end{figure*}

For \astrobj{ALS 3721}, we calculate the theoretical amplitude ratios of the $0\le l \le3$ modes for the models assuming a mass range between $10M_\odot \le M \le 14M_\odot$ and  effective temperatures in the range of  $23000 < T_{\rm eff} < 25000$~K.  The derivation of the theoretical values of different $l$ degrees in a certain range of  $T_{\rm eff}$ and $\log g$ were obtained by using $1.6 < M_\odot < 20$, ZAMS to TAMS and A04, {\tt OPAL}\footnote{{\tt OPAL} opacities computed with non-adiabatic Warsaw-New Jersey/Dziembowski code by J. Daszy{\'n}ska-Daszkiewicz  and P. Walczak ({\it http://helas.astro.uni.wroc.pl}).}. The theoretical expression of the model (with the assumption of zero-rotation and static plane-parallel atmospheres) included in the software package was represented by \citep{daz02, daz03}. The precomputed grids of the atmospheric parameters have been calculated by applying {\tt Kurucz} and {\tt NEMO} atmospheres and the quality of the solution is represented by $\chi^2$ \citep{zim08}. As shown in Fig.~\ref{figm1}, the amplitude ratios were plotted as a function of the central wavelength normalized to the U filter for the highest amplitude frequencies $f_{1}$=4.9457 and $f_{2}$=5.3928. For the primary pulsation frequency, $f_1$, the relation between the predicted values and the observed ones are not too much distinctive for the masses of $10M_\odot$ and $14M_\odot$ . It may be concluded that the amplitude ratios are intermediate between $l=0$ and $l=1$. In contrary, the masses between the values of  $10M_\odot$ and  $14M_\odot$ are remarkably represented with the curve of $l=0$.  The calculated amplitude ratios for $f_2$ except for the mass of $14M_\odot$ are matched with the observations and lead to value of $l=1$
but the results are not so clear as those of masses for  $14M_\odot$. Consequently, mode identification is inconclusive  for this mass and it could be considered between $l=0$ and $l=1$ .

\section{Discussion}
\astrobj{ALS 3721} was confirmed to be possessed $\beta$ Cephei type pulsation characteristics. Frequency analysis based on photometric observations resulted with the two reasonable frequencies, $f_{1}$=4.9457 and $f_{2}$=5.3928, which are previously proved by \cite{pig08}. Spectroscopic observations and analysis indicate that \astrobj{ALS 3721} has a high rotational velocity ( $v \sin i$=$100 \pm 20$\,km\,s$^{-1}$) with derived effective temperature $T_{\rm eff} = 24000\pm1000$\,K and $\log g = 3.5\pm0.1$. Additionally, the luminosity of the star is found to be log${L \over L_{\odot}}$=3.22 based on the calculation from the distance modulus of \astrobj{NGC 6200}, $m-M=13^{m}.36$, given by {\tt WEBDA}\footnote{{\it http://www.univie.ac.at/webda/webda.html}} database. During the calculation, the values of apparent magnitude of the star and the solar absolute magnitude was adopted from \cite{pig08} and \cite{cox00}, respectively. \astrobj{ALS 3721} is a member of the Southern open cluster \astrobj{NGC 6200} and it is an early age cluster.  In this case, if all results were taken into consideration it might be concluded that the star was located on the ZAMS and still at the stage of core Hydrogen burning in the instability strip.

One of the bottlenecks for existing mode identification methods is to determine the order of spherical quantum numbers in an accurate way. Recently, with new generation space missions like {\it Kepler} \citep{bor10}, {\it CoRoT} \citep{bag06} and {\it MOST} \citep{wal03} stellar variability is re-identified by analyzing unprecedented data sets which makes possible the detection of the very low amplitude oscillations with a high precision level. Alas, current space based data are not compatible with existing models including both spectroscopic and photometric methods and thus astreoseismic studies are still in need of the ground based data. From the ground based observations, $\beta$ Cephei stars are distinguished with typical frequencies in the high frequency domain. Considering that, these stars are relatively more convenient targets in order to test the reliability of the stellar models. Since it is assumed that they have no deep surface convection zones, several physical conditions would be negligible in the model computations. Because of this, seismic interpretation of the pulsation modes can be more possible for at least a few frequencies, however, this is not a satisfying solution in some cases such as rapid rotation. What is more, model computations are more complicated than expected in the presence of rapid rotation.  In other words, if the star rotates very fast, the effects of rotation on eigenfrequencies should be additionally considered for a trustworthy seismic modelling. For the $\beta$ Cephei stars , another important effect on the detection limits of the oscillation modes were also reported by \cite{daz02}. They have found that the determination of higher $l$ modes ($l>2$) only from multicolour photometric data may not be expected. Consequently, such $l$ modes may be expectable in highly densed frequency spectra obtained with space based instrumentation. On the other hand, the detection of intermediate $l$ modes  ($l \le 2$) are more likely possible by using spectroscopic line profile variations and photometric amplitude ratios.
In our study, the spherical harmonic degrees have been computed for the two of strongest frequencies of \astrobj{ALS 3721}. In addition to the photometric data, spectroscopic data were also achieved to extract fundamental stellar parameters. For mode identification, we used   {\tt FAMIAS}  software package \citep{zim08} to obtain amplitude ratios corresponding to $l$  degrees assuming the masses between $10M_\odot \le M \le 14M_\odot$. Model outcome for the mass of $10M_\odot$ and $14M_\odot$ as seen in Fig.~\ref{figm1}, the distinction between $l=0$ and $l=1$ modes for $f_1$ is not clear in the amplitude ratio. The results for $f_2$ the amplitude ratios are consistent for a mass range of between  $10M_\odot \le M \le 13M_\odot$ and reproduced by the curve for $l=1$. Nevertheless, observed and calculated colour amplitude ratios are not in a best agreement each other for the mass of $14M_\odot$  and mode identification is ambiguous between $l=0$ and $l=1$.

Taken together, we found that the amplitude ratios of \astrobj{ALS 3721} are between $l=0$ and $l=1$ degrees assuming masses in the range of $9M_\odot$ and $15M_\odot$ which coincides $\beta$ Cephei instability domain on the HRD. As we mentioned above, from the ground based data, we can only determined the mode parity of the $l$ degrees in the range of intermediate modes. In general, therefore, our results seem in a  very good agreement with the theoretical expectations.

\section*{Acknowledgments}
CU wishes to thank the South African National Research Foundation for the prize of innovation post doctoral fellowship with the grant number 73446. EN acknowledges support from the NCN grant 2011/01/B/ST9/05448. The authors wish to thank SALT astronomer Dr. Petrie Vaisanen for helping the spectroscopic observations.The authors made use of the {\tt WEBDA} database, operated at the Institute for Astronomy of the University of Vienna, and of the NASA Astrophysics Data System.

\end{document}